\documentclass[sigconf]{acmart}
\AtBeginDocument{%
  }

\copyrightyear{2025}
\acmYear{2025}
\acmConference[VRCAI '25]{The 20th ACM SIGGRAPH International Conference
on Virtual-Reality Continuum and its Applications in Industry}{December
13--14, 2025}{Macau, China}
\acmBooktitle{The 20th ACM SIGGRAPH International Conference on
Virtual-Reality Continuum and its Applications in Industry (VRCAI '25), December
13--14, 2025, Macau, China}\acmDOI{10.1145/3779232.3779278}
\acmISBN{979-8-4007-2362-9/2025/12}

\begin{document}

\title{Epistemoverse: Toward an AI-Driven Knowledge Metaverse for Intellectual Heritage Preservation}

\author{Predrag K. Nikoli\'{c}}
\orcid{0000-0003-1409-088X}
\affiliation{%
  \institution{School of Design and Arts\\Swinburne University of Technology}
  \city{Kuching}
  \country{Malaysia}
}
\email{pnikolic@swin.edu.au}

\author{Robert Prentner}
\orcid{0000-0003-1890-0827}
\affiliation{%
  \institution{Institute of Humanities\\ ShanghaiTech University}
  \city{Shanghai}
  \country{China}
 }
\email{robert.prentner@amcs.science}

\renewcommand{\shortauthors}{Nikolic \& Prentner}

\begin{abstract}
  Large language models (LLMs) have often been characterized as “stochastic parrots” that merely reproduce fragments of their training data. This study challenges that assumption by demonstrating that, when placed in an appropriate dialogical context, LLMs can develop emergent conceptual structures and exhibit interaction‑driven (re-)structuring of cognitive interfaces and reflective question-asking. Drawing on the biological principle of cloning and Socrates’ maieutic method, we analyze authentic philosophical debates generated among AI-reincarnated philosophers within the interactive art installations of the Syntropic Counterpoints project. 

By engaging digital counterparts of Aristotle, Nietzsche, Machiavelli, and Sun Tzu in iterative discourse, the study reveals how machine dialogue can give rise to inferential coherence, reflective questioning, and creative synthesis. Based on these findings, we propose the concept of the Epistemoverse—a metaverse of knowledge where human and machine cognition intersect to preserve, reinterpret, and extend intellectual heritage through AI-driven interaction. This framework positions virtual and immersive environments as new spaces for epistemic exchange, digital heritage, and collaborative creativity.
\end{abstract}


%

\maketitle

\section{Introduction}
In biology, a clone is defined as an individual that is genetically identical to the original organism from which the clone was derived \cite{Raff2018}. Throughout biology, cloning is a ubiquitous reproductive strategy that can be observed in a variety of species, mostly in bacteria and plants. It is essential to emphasize that a clone differs from a copy. Whereas a clone is genetically identical to the organism it is derived from, it can potentially undergo an individual developmental trajectory. Virtually all biological traits, such as cognitive differences among individuals, arise as an intricate combination of genetic predisposition and environmental facilitation \cite{BouchardMcGue2003}.

This is different from the way we typically think about software agents, which are mere copies of a program instantiated on different computers and executed in a well-defined and identical manner. In this contribution, we are investigating AI clones of historically existing philosophers \cite{NikolicLiuLuo2021}. We aim to demonstrate that those clones, while implemented via fine-tuning an LLM \cite{FierroEtAl2024} on the corpus of their “ancestors” (Aristotle, Nietzsche, Machiavelli, and Sun Tzu, respectively), exhibit remarkable kinds of knowledge that go beyond the knowledge that we would have uncovered if we studied them in isolation.

Analysis should focus on the intrinsic perspective of philosophers, not on the correspondence between their utterances and human knowledge. We are not interested in how those clones would fare against human philosophers or whether they are just repeating what those philosophers have said (unlike what was presented in recent lawsuits against AI companies, who allegedly violated the copyrights of real human authors). By contrast, we are mainly interested in whether our AI clones could produce knowledge that stems from interaction-induced conceptual organization, beyond isolated reproduction. Put more provocatively, their knowledge might be said to be ``genuinely theirs.'' If this is the case, there is a hard limit beyond which we may not recognize it. Thus, this paper contributes to making the conceptual structures emergent from LLMs more transparent and interpretable.

From this inquiry arises a broader framework we refer to as the \textit{Epistemoverse}: an evolving digital ecosystem in which AI-driven philosophical agents not only reproduce but also extend human intellectual heritage through interactive reasoning. Within this framework, knowledge becomes a dynamic and relational process, an interplay between human interpretation and machine cognition that can eventually unfold within immersive, interconnected environments. It is essential to emphasize that, ultimately, we aim for these environments to be as complex and immersive as possible, potentially utilizing VR/AR technology. From a perspective based on the idea of a “clone” (not a mere copy), this is not optional but mandatory. 

We will first provide background on the art project \textit{Syntropic Counterpoints}, starting with the art installation \textit{Robosophy Philosophy} and then moving on to the \textit{Botorikko} and \textit{Metaphysics of the Machines} artworks. We then introduce the technical approaches used to generate philosophical discussions between AI clones, ranging from early RNN models to more advanced multi-agent dialogue systems. Our methodology for analyzing the AI-generated philosophical debates will follow, focusing on extracting networks from machine-generated text and examining their structure. Finally, we will discuss the implications of our results for understanding machine creativity and cognition, as well as their applications in the humanities.

Following the interface-theoretic view of cognition \cite{PrentnerHoffman2024,Prentner25ac}, we treat our AI philosopher-clones as cultural interfaces that mediate access to a relational substrate of philosophical knowledge rather than as inner models with hidden mental states. On this view, “knowledge” is expressed in stable patterns of interaction and constraint among agents and sources, not in isolated propositions. The \textit{Epistemoverse} is the mesoscale where such interface-level regularities become observable in multi-agent discourse. Methodologically, our analyses target this interface organization (who connects to what and when) and its condition-dependence (with/without maieutic links), rather than claims about internal cognition.

\section{Background}
Starting in 2017, the \textit{Syntropic Counterpoints} project explores AI as a creative medium and the potential of using robot-robot and human-robot interaction to highlight new phenomena that are interesting to examine through an interactive media art prism for further critical observation. To raise relevant questions within the humanities, we generate philosophical discussions between AI clones of renowned philosophers from different eras and cultural origins \cite{NikolicLiu2021,NikolicTomari2021}. This approach challenges traditional AI usage and investigates the concepts of AI abstraction, creativity, and consciousness \cite{NikolicTomari2021}. Our focus in this paper will be on authentic philosophical debate analysis generated by the clones and their evolution from RNN to LLM neural network models, applied as they were introduced for further usage. 

\subsection{Robosophy Philosophy}
This installation, titled \textit{Syntropic Counterpoints: Robosophy Philosophy}, was envisioned as a series of philosophical discussions conducted between AI clones of Aristotle and Nietzsche. These virtual thinkers were engaged in debates on topics such as morality, ethics, and aesthetics. Our aim was to subject these cloned philosophers to their own interpretations of human knowledge, and point out ongoing cultural and social shifts resulting from human-technology interactions \cite{Nikolic2020}.

\begin{figure}[hbt]
\centering
\includegraphics[width=0.48\textwidth]{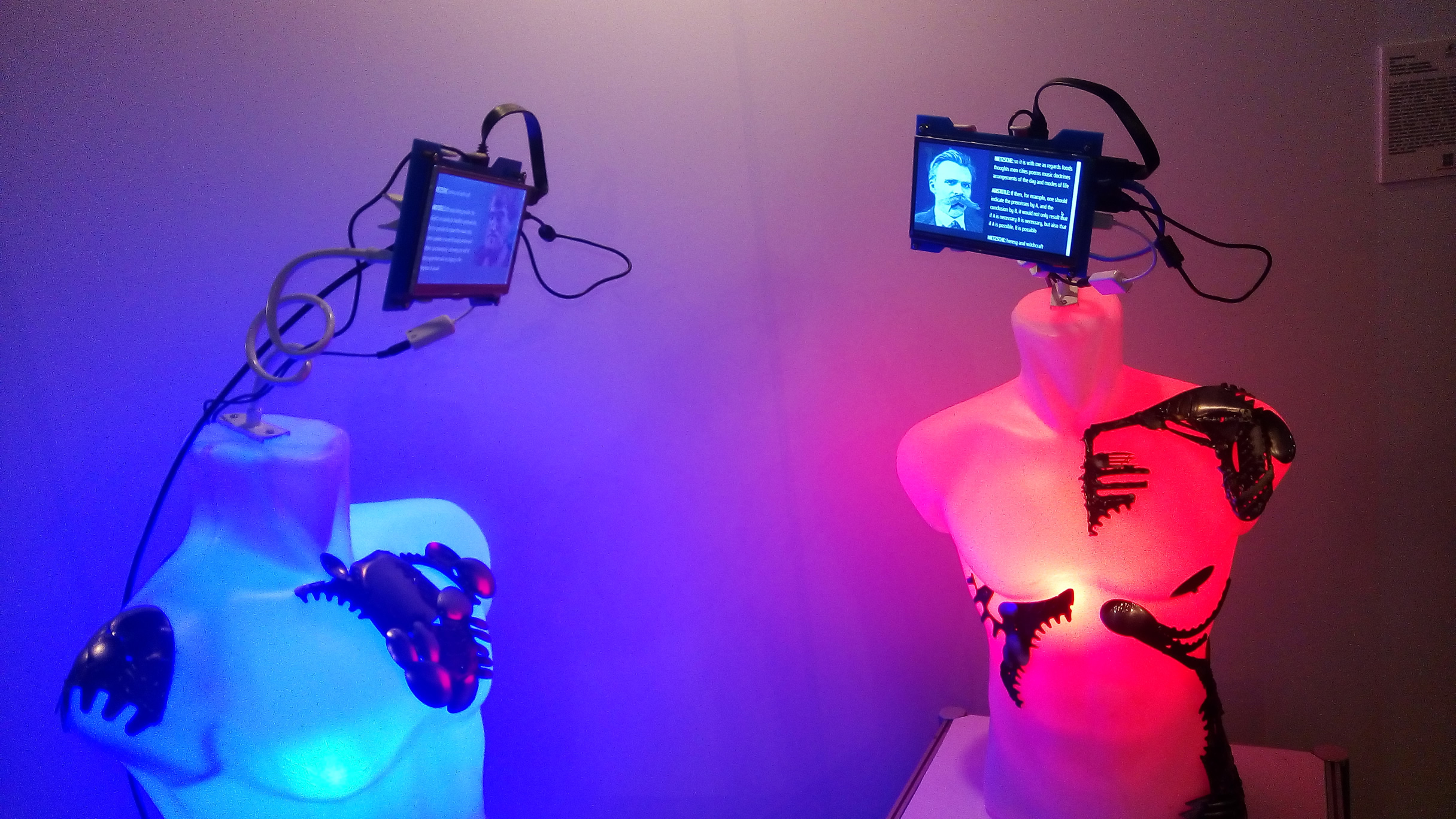}
\caption{\textit{Robosophy Philosophy}: AI-cloned Aristotle and Nietzsche engage in a techno-cloned dialogue exploring moral, ethics, and virtues.}
\label{fig:1}
\end{figure}

Our objective was to unite two philosophers in a timeless dialogue (Figure \ref{fig:1}), through the use of AI clones and modern technology (“techno-cloning”) \cite{NikolicEtAl2018}.

The resulting debates were authentic creations of AI agents, providing a robust foundation for scrutinizing the employed neural network model (RNN) through the lens of AI language usage, grammar, and contextual comprehension of the presented knowledge corpus \cite{Nikolic2020}.

\subsection{Botorikko, Machine-Created State}
\begin{figure}[hbt]
\centering
\includegraphics[width=0.48\textwidth]{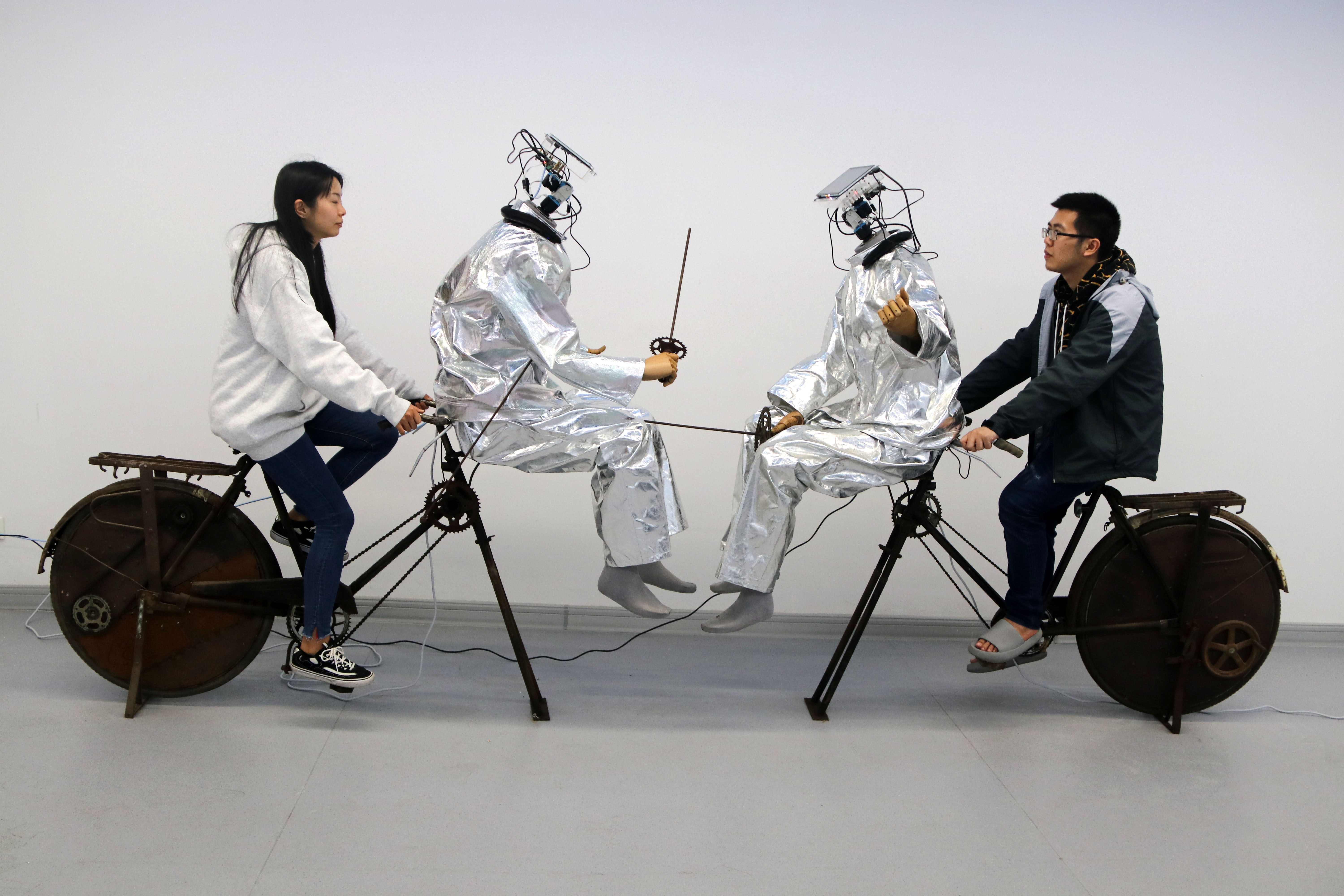}
\caption{\textit{Botorikko, Machine-Created State}: AI clones of Machiavelli and Sun Tzu engage in real-time debates on politics, diplomacy, strategy and conflict.}
\label{fig:2}
\end{figure}

For the subsequent interactive installation, \textit{Syntropic Counterpoints: Botorikko, Machine-Created State}, we techno-cloned Machiavelli and Sun Tzu. The two bicycles transported pseudo-robotic mannequins with monitor-like heads. Visitors can listen to the speakers and observe real-time dialogues displayed on computer displays. We subject our artificial intelligence clones to information pertaining to politics, diplomacy, warfare, and human conflicts \cite{NikolicTomari2021}. We employed the metaphor of sword fighting triggered with pedaling engagement to engage visitors in an absurd interaction open to various interpretations (Figure \ref{fig:2}).

The movement of the robot's head, in response to the sentiment of machine-generated content, plays a crucial part in the conceptual and aesthetic framework of the \textit{Botorikko, Machine-Created State} artwork. It is wholly produced using artificial intelligence's analysis of its conversational content \cite{NikolicTomariJovanovic2021}.

\subsection{Metaphysics of the Machines}
\begin{figure}[hbt]
\centering
\includegraphics[width=0.48\textwidth]{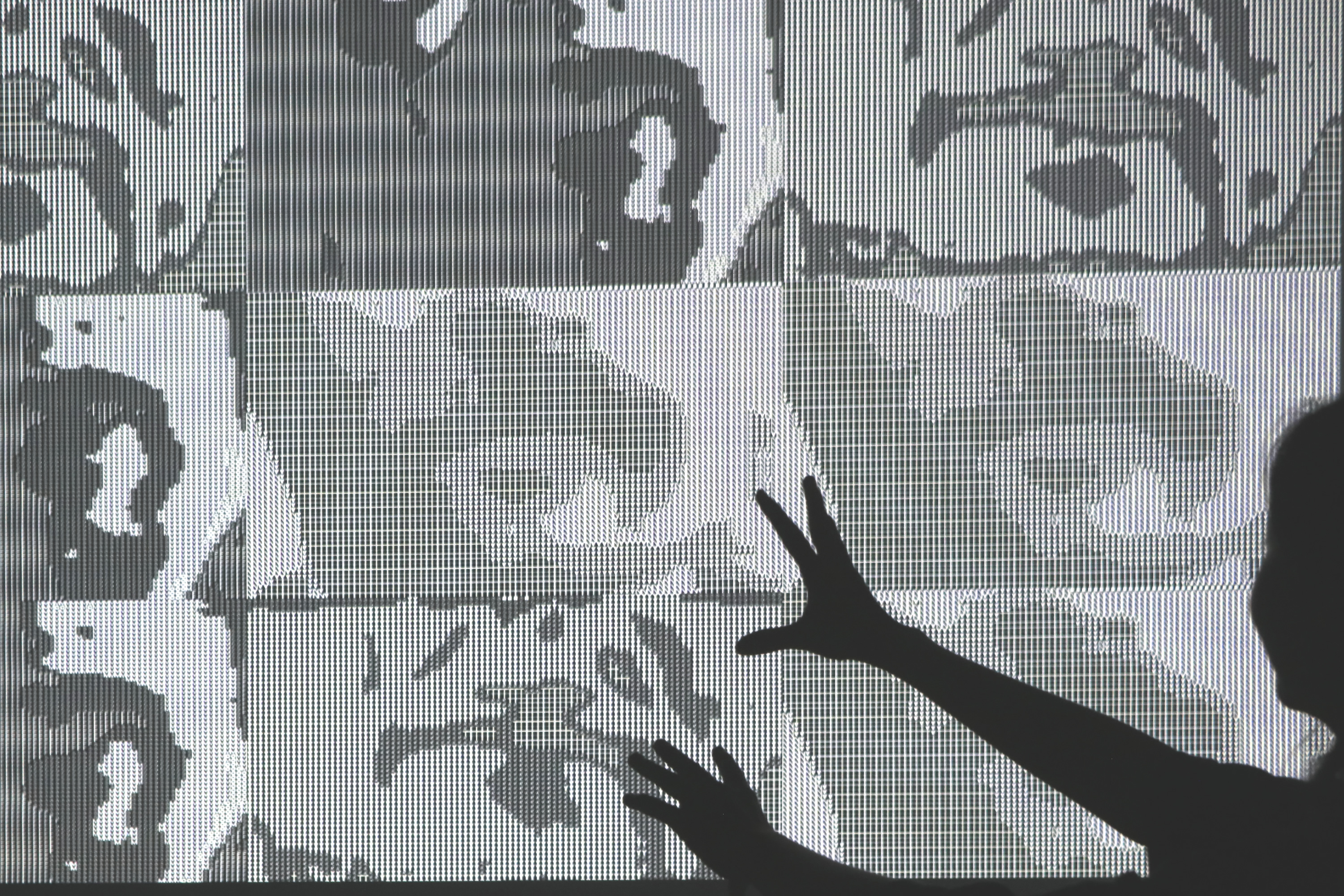}
\caption{\textit{Metaphysics of the Machines}: AI-cloned philosophers Aristotle, Nietzsche, Sun Tzu, and Machiavelli engage in autonomous debates on metaphysics, morality, and existence, generating self-evolving dialogues that question machine reasoning and abstraction}
\label{fig:3}
\end{figure}

Our aim in the art installation \textit{Syntropic Counterpoints: Metaphysics of the Machines} was to further investigate AI aesthetic phenomena and challenge the abstraction of machines. We subjected philosophers’ AI Clones to essential philosophical inquiries regarding the connection between mind and existence, as well as metaphysics. To achieve this objective, we developed four AI Clones: Aristotle, Nietzsche, Sun Tzu, and Machiavelli, each trained on significant philosophical discourses bequeathed by these thinkers to humanity \cite{NikolicLiuLuo2021}. We are initiating their discourse with eternal questions such as “Why is there something rather than nothing?”, "Is war moral and ethical, and can it ever be justifiable?" or "What is good and what is evil?" Subsequently, relinquish the discourse entirely to the AI cloned philosophers to advance it further, formulating their inquiries based on the generated content and addressing additional subjects such as progress, law, morality, and other esteemed qualities that numerous intellectuals have explored in various texts throughout human history. The philosophers' AI clones adhered to distinct patterns and employed inventive vocabulary derived from the world, grammar, and letters to present an additional alternative reality. It transcends a realm constructed from calculations, algorithms, projections, probabilities, and automated decisions that we anticipate will elucidate life's enigmas, provide solutions to humanity's enduring questions, and determine our choices. The installation “Metaphysics of the Machines” employs irony and skepticism to critique the human ability to comprehend works that transcend logic and pragmatism, as well as realms beyond human perception (Figure \ref{fig:3}).

Our Artificial Intelligence System Philosopher Clones possess full authority over the execution and manifestation of their philosophical discourse within an alternative world displayed on the wall. We focus on language, specifically how AI clones process it. The outcome remains uncertain, unknown until it occurs, thereby challenging the entire notion of knowledge \cite{NikolicLiu2021}.

\section{Generating AI Clones Discussions}
For the analysis presented in this paper, we utilized a multi-agent dialogue system \cite{PapangelisEtAl2019,PapangelisEtAl2020} to generate discussions among AI agents with distinct knowledge and domains in the interactive installation "Metaphysics of the Machines." Each agent utilized a neural language generator to generate plausible responses conditioned on input queries. Addressing the challenge of limited data for historical clones, the system adopted a GAN-based architecture with a retriever-discriminator mechanism \cite{LewisEtAl2020}. The retriever utilized a pre-trained BERT encoder to map books 
into dense representations for knowledge retrieval. A shared seq2seq model \cite{StrobeltEtAl2019} language generator, fine-tuned on individual texts, facilitated dialogue.

Implementation details and reproducibility. Each AI clone operates a philosopher‑specific retriever coupled to a shared generator fine‑tuned per corpus via parameter‑efficient updates. Corpora comprised Aristotle's \textit{Nicomachean Ethics, Poetics, Politics}, and \textit{Metaphysics}, Nietzsche's \textit{Thus Spoke Zarathustra, The Antichrist, Beyond Good and Evil, The Gay Science, The Birth of Tragedy}, and \textit{Ecce Homo}, Machiavelli's \textit{The Prince}, and Sun Tz's \textit{The Art of War}. The retriever architecture comprised a query encoder, document encoder (using pre-trained BERT; \cite{Devlin2019}), and probabilistic retriever. For each input query, the system identified top-k document chunks that are most likely to provide accurate answers to the posed question, represented mathematically as:
\begin{equation}
p_i(z|x) \propto \exp\left( d(z),q(x) \right),\ i \in [1,k].
\end{equation}
The communication process among the four robots was conceptualized as a retrieval-based generation process. The retrieved component $p_i(z|x)$ was subsequently marginalized to form a probability distribution over a pre-trained seq2seq vocabulary, which is integrated with the generator component. The generator component was derived from a latent code generated by an encoder-decoder architecture, based on a fine-tuned version of GPT-2.  

\begin{figure}[hbt]
\centering
\includegraphics[width=0.48\textwidth]{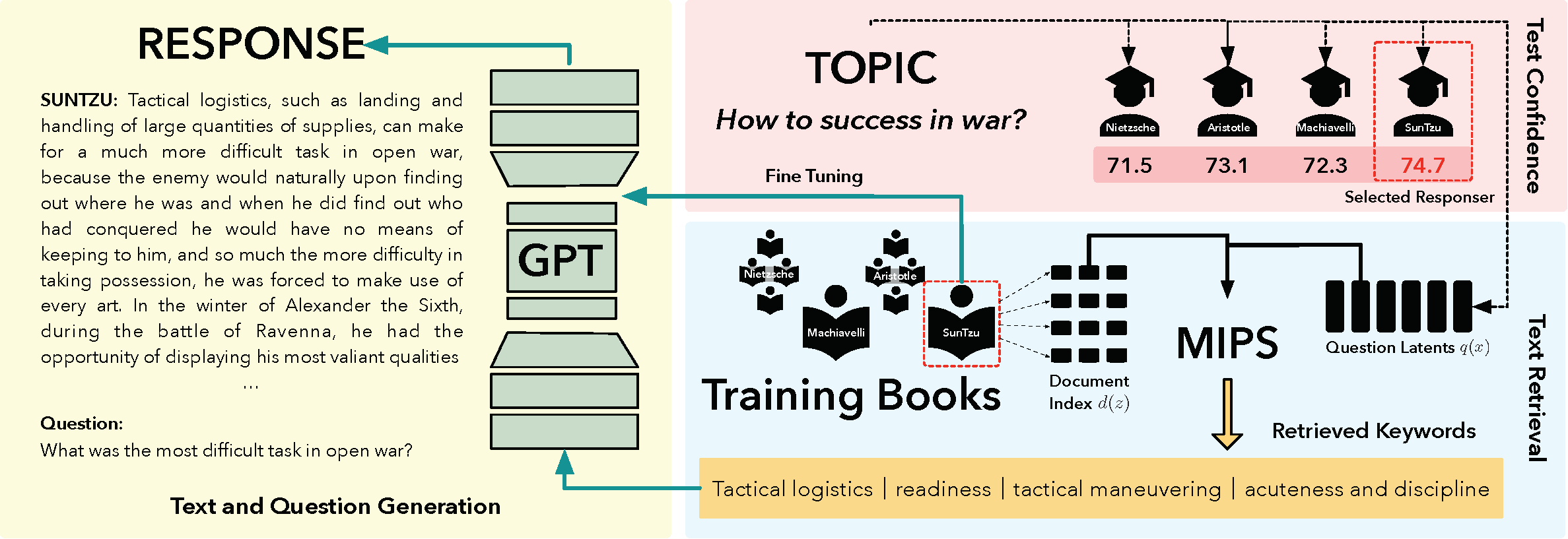}
\caption{This diagram corresponds to our RAG pipeline: philosopher‑specific retrievers rank candidate chunks, a shared generator fine‑tuned per corpus produces responses, and MIPS selects the next speaker (see Implementation details).}
\label{fig:4}
\end{figure}

In initial experiments, a randomly selected topic initiated debates. Robots retrieved relevant knowledge chunks, and responses were generated with fine-tuned GPT models. Confidence in responses was measured using the Maximum Inner Product Search (MIPS) algorithm. In subsequent experiments, robots formulated their own questions, encouraging dynamic debates. For instance, Machiavelli might ask, “What was the most difficult task in open war?” while Sun Tzu inquired, “What does heaven mean?” (Figure \ref{fig:4}).

The final phase introduced complex philosophical questions like “Do we make war that we may live in peace?” and “Where is the line between art and not art?” These questions, drawn from the writings of the cloned philosophers, provided a foundation for intricate discussions. Adjusting keyword relevance and retrieval parameters further enhanced the debates’ depth and variety, pushing the boundaries of AI-generated philosophical discussions.

\section{Research Approach}
A first clue to the idea that AI philosophers are more than mere stochastic parrots is suggested by the philosophers asking questions rather than merely answering humans. Typically, LLMs are used to provide answers to well-defined prompts rather than to generate insightful questions themselves. The latter requires either a reflective capacity or the presence of a knowledgeable teacher who could direct them accordingly. Both are cornerstones in the way we teach philosophy students today. In the history of philosophy, this technique is well-known as Socrates’ maieutic method \cite{Candiotto2015}. 

Socrates believed that this would help philosophizing subjects to “remember” a variety of inborn and ideal concepts. In the context of AI clones, this could be interpreted in two ways. The maieutic method could be thought of as leading to the uncovering of implicit knowledge that is hidden in the corpus and needs to be “made explicit” via prompts or it could lead to the emergence of entirely new kinds of knowledge. Either way, the method produces a deeper understanding and a higher state of reflection than before the conversation. 

We operationalize the “knowledge” of our AI philosophers via the notion of “concept networks.” \cite{MargolisLaurence2023} A concept network is defined via the relations between concepts extracted from text. It can be visualized as a graph, where nodes represent concepts and edges represent the links between them. In our analogy with biological clones, graph nodes are represented by keywords extracted from generated text, standing for the (expression) of genetic material from human ancestors. In practical terms, these concepts can be derived from the answers of our AI clones to specific questions they have been prompted with. For example, taken from a conversation recorded during the “Metaphysics of the Machines” installation, an AI clone based on the philosopher Friedrich Nietzsche was asked:
\\\\
``What is the meaning of inequality and being unequal?''
\\\\
to which the clone replied:
\\\\
``[It] is, as it were, a form of deadly hostility to everything on earth. On the other hand, the Gospels were indeed read as the book [of] innocence, but this is entirely to overlook the fact that they were read as the book of malice. The Gospels were read as the book of hatred, which [is a] small indication of the ultimate artistic with which this piece of dramas been played out. The instinct of resentment in favour of life was the first to take seriously that tremendous possibility that the small congregation might be the true readers [of] the Bible: and that in fact, we might one day find it even more necessary than the closing sections of the first book. His possibility is close to what I called the security concern of the Evaluation, a prospect that has not been expressed [at] all because I was too serious about the truth of the matter.''\\ 

We used the Python library NLTK to extract five keywords that reflect the main motifs expressed by the AI philosopher. In this particular case, the keywords were: 
\begin{equation}
\textrm{\textbf{book, read, gospel, fact, possibility}}.
\label{eq:i}
\end{equation}
The recorded conversations were structured along many such fragments (“chunks”), such as the one given above, usually in response to an artificially generated question. Chunks are assumed to encode a single “concept” held by the AI philosopher (in the graph, individual concepts correspond to fully connected sets of five keywords). The concepts derive from the textual material on which those clones have been trained. Going through all the chunks, it is possible to extract all the nodes of the clone’s concept network, where every node corresponds to a list of five keywords. However, knowledge is more than an unstructured bag of concepts. Knowledge also needs structure. This structure is determined by the relations between concepts. In a graph, these relations are represented as edges. Edges reflect either (1) lexical overlap (shared keywords), or (2) a maieutic link (question‑mediated adjacency across agents). This heuristic intentionally favors transparency over exhaustiveness. We acknowledge that surface overlap can miss paraphrastic links and that not all overlaps indicate deep semantic ties; our conclusions are therefore framed in terms of an interaction‑driven (re-)structuring of cognitive interfaces rather than claims about the ``true nature'' of conceptual representation.

\begin{enumerate}
\item One possibility is that the edges represent conceptual similarity or overlap between individual concepts. For example, a concept expressed much later in the dialogue is represented by the keywords
\begin{equation}
\textrm{\textbf{principle, hostility, amount, friendship, fact}}.
\label{eq:ii}
\end{equation}
It does overlap with \eqref{eq:i} (i.e., there is one shared keyword, namely “fact”). The following lists of keywords overlaps both with \eqref{eq:i} and \eqref{eq:ii} (again in the keyword ``fact''):
\begin{equation}
\textrm{\textbf{philosophy, fact, sit, thing, people}}.
\label{eq:iii}
\end{equation}
Such a list of partly overlapping keywords gives rise to a “connected component” within the graph. Note there are several keywords in the concept network without having links to any keywords outside their chunk (thus not forming a connected component with other such keywords), for example 
\begin{equation}
\textrm{\textbf{master, morality, reality, mean, modernity}}.
\label{eq:v}
\end{equation}
The concept network of the AI clone of Nietzsche is shown in Figure \ref{fig:5} (shown are only the first 9 chunks).   
 
\begin{figure}[hbt]
\centering
\includegraphics[width=0.48\textwidth]{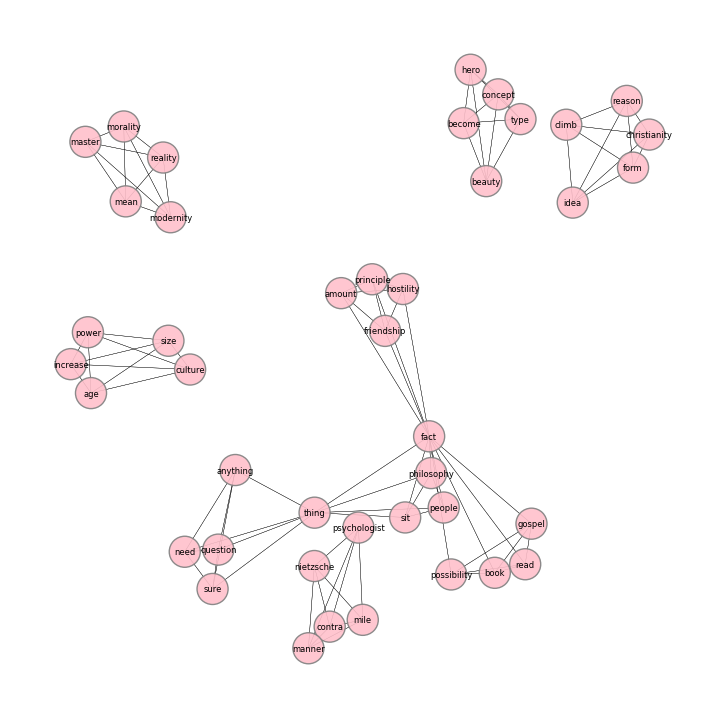}
\caption{Concept network of the AI clone of Nietzsche. A “connected component” is formed by several keywords from different chunks of text, including the keywords “fact,” “thing,” and “morality” that are part of different chunks.}
\label{fig:5}
\end{figure}

\item Another possibility, representing the “reflection” of AI, is that such edges correspond to connections between concepts that are mediated by questions of others. For example, the keywords \eqref{eq:i} would trigger the question,
\\\\
“What do you think [of] the Gospels?”,
\\\\
which is sent to the AI clone of Aristotle, who in turn asks back the question,
\\\\
“What is the difference between voluntary and involuntary actions?”
\\\\
which then leads to a new concept, represented by the keywords:

\begin{equation}
\textrm{\textbf{question, need, anything, thing, sure}}
\label{eq:iv}
\end{equation}

As one can see, there is no overlap in keywords between \eqref{eq:i} and \eqref{eq:iv}, but they are still indirectly connected via an intermediate question. We call such connections, which have been triggered by the interlocutions of other AI clones, a ``maieutic link.'' 
\end{enumerate}
Therefore, edges can correspond to either genetically encoded links between nodes or developmentally acquired connections. In the first case,
the knowledge in question could be called “inborn” (genetically pre-determined); in the latter case, knowledge could be said to be “emergent.” Figure 6 shows the concept networks of all four AI clones that participated in the discussion triggered by the question:

“Is war moral and ethical, and can it ever be justifiable?”

AI clones of Aristotle and Machiavelli tend to form one large connected component each. By contrast, the concepts of AI clones of Machiavelli and Sun Tzu stay relatively isolated. If one zoomed in onto the nodes, one would find a fine-structure corresponding to the lists of 5 fully connected keywords shown in Figure \ref{fig:5}. It is worthwhile to stress at this point that the initial question was the only human input given to the AI philosophers’ discussion. The remaining questions throughout the discussion were entirely machine-created.
 
\begin{figure}[hbt]
\centering
\includegraphics[width=0.48\textwidth]{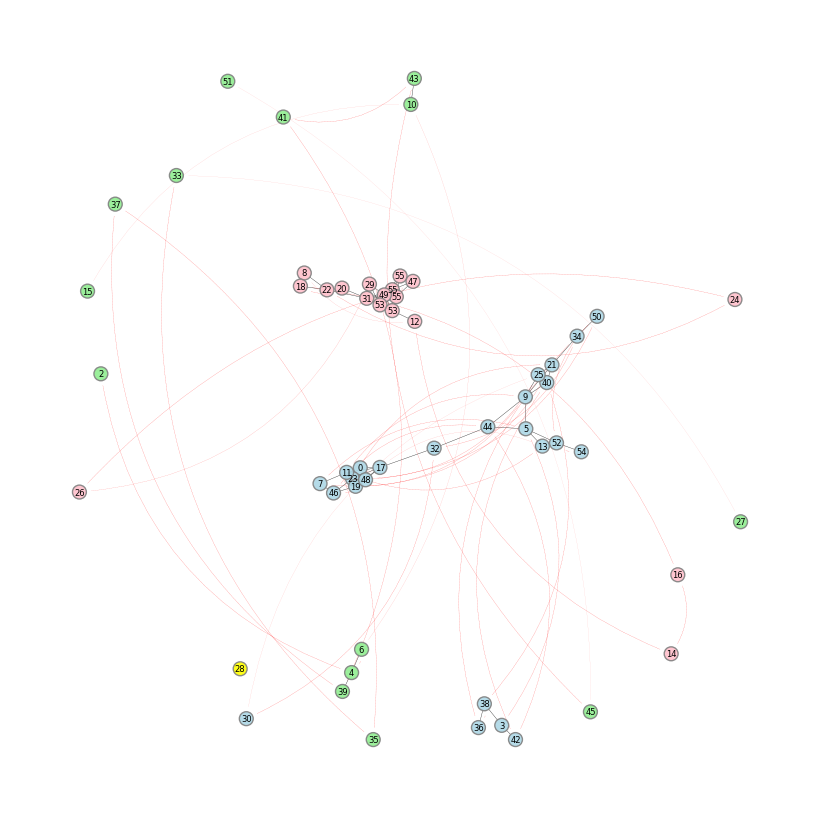}
\caption{Concept networks of AI clones triggered by the question, “Is war moral and ethical, and can it ever be justifiable?” The color code is as follows: “blue” stands for Aristotle, “red” for Nietzsche, “green” for Machiavelli, and “yellow” for Sun Tzu. Each node in the graph represents a concept from a discussion part of “Metaphysics of the Machines,” labeled by an “index” that numbers any chunk of text in the dialogue. The total number of chunks included is ca. 60. Links in black represent the clone’s “inborn” knowledge. By contrast, links in red are “maieutic links” that stand for the emergent knowledge required throughout the dialogue.}
\label{fig:6}
\end{figure}

Concept networks are read here as phenomenological approximations \cite{Prentner2025} of the interface: they summarize how discourse affords and constrains further moves across agents. We model regularities in representation at the interface, not ``inner'' representational states. Degree-based measures and edge types (lexical vs. maieutic) are therefore proxies for interaction-driven organization: how the interface channels and stabilizes exchanges, rather than truth-theoretic or cognitive attributions. This could be analyzed using well-studied measures from graph theory \cite{Newman2010}. For example, a widely used measure would be the average degree of centrality $C_d$ of a graph. For a graph with n nodes $v \in V$, any node could have $n-1$ links (not counting self-loops). The degree of a node is then defined as the number of connections normalized by this number. One could then average this over the nodes in the graph, thus giving:

\begin{equation}
C_d=\frac{1}{n} \sum_{v \in V} \textrm{deg}(v).
\end{equation}

In our analysis, which dates back to the initial installation of \textit{Robosophy Philosophy}, we focus on the conceptual structures of Aristotle and Nietzsche, as their respective corpora of works were sufficiently large for AI clones to be trained on. By contrast, Machiavelli and Sun Tzu clones, the featured thinkers in the \textit{Botorikko, Machine-Created State} installation, primarily acted as facilitators of maieutic connections. 

\begin{figure}[hbt]
\centering
\includegraphics[width=0.48\textwidth]{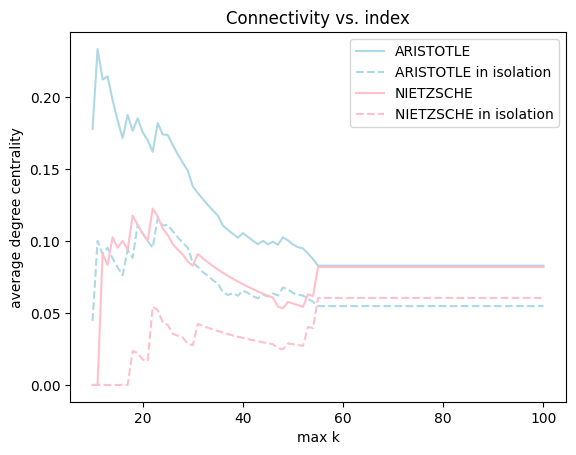}
\caption{Average degree centrality, illustrating the graph's connectivity as a function of the number of indices (chunks) in a philosophical dialogue. After ca. 60 chunks, the degree centrality converges. Shown are curves for AI clones of Aristotle and Nietzsche (with and without interaction with other philosophers). Not shown are the plots for Machiavelli and Sun Tzu, but their presence significantly increased the average degree centrality through maieutic links.}
\label{fig:7}
\end{figure}

Reported results are thus for Aristotle and Nietzsche. We can calculate the average degree centrality as a function of the number of question/answer pairs (“indices” $k$) throughout the dialogue. Figure \ref{fig:7} shows this function for the clones of Aristotle and Nietzsche, with and without connections via maieutic links.

We report two conditions: (i) maieutic (cross‑agent question‑asking enabled), and (ii) question‑off (no cross‑agent questions; retrieval and generation unchanged). Figure \ref{fig:7} contrasts these conditions for Aristotle and Nietzsche as a function of dialogue indices $k$. Qualitatively, the maieutic condition yields denser graphs (higher average degree centrality) and fewer isolates, which suggest an interaction-driven (re-)structuring of the agent's cognitive interface.

To complement degree centrality, one could compute embedding\-based local coherence, for example, by averaging cosine similarity of adjacent fragments. These lightweight measures would align with our graph‑level findings while avoiding direct claims about reasoning. The analysis suggests that including the “pingback” facilitated by asking questions to other clones increases a basic measure of connectivity in the individual clones' concept network. Their knowledge becomes “denser,” independent of human intervention. Interestingly, the results appear to be independent from the only human-made input, which was the initial question. 

\section{Conclusion and Future Directions}
This contribution demonstrates the productivity of combining methods from AI research, art, and computational analysis to generate and examine hybrid human–machine artifacts (“techno-cloning”). Our analyses, based on the \textit{Syntropic Counterpoints} installations, offered an initial exploration of emergent behaviors in Human-Robot-Robot interaction. By analyzing artificially generated dialogues among AI clones of historical philosophers, we revealed that interaction between these agents produced distinctive conceptual networks. When comparing the clones' concept graphs in isolation with those that emerged from their dialogues, we observed measurable increases in graph-theoretical metrics, particularly degree centrality, indicating a rise in semantic connectedness, conceptual interdependence, and the emergence of new inferential links through interaction-driven organization as reflected in graph connectivity.

While the \textit{Metaphysics of the Machines} installation already extended dialogue generation using GPT-based architectures, future iterations should further refine analytical precision by integrating transformer-based language models into the interpretive process itself. Rather than relying exclusively on statistical measures from NLTK, these models could assist in context-sensitive keyword extraction, topic segmentation, and the semantic clustering of philosophical constructs across dialogues.

This study marks only the first step toward a more expansive framework for artificial sense-making. We anticipate that sustained interactions between artificial agents, humans, and socio-technical environments will continue to generate novel epistemic behaviors, instances where AI systems contribute not only to discourse but to the co-production of meaning. To capture and extend this phenomenon, we introduce the conceptual model of the Epistemoverse: a dynamic knowledge metaverse that situates AI philosopher clones, human interlocutors, and analytical systems within a shared epistemic space.

In the \textit{Epistemoverse}, techno-cloned agents function as living knowledge entities, continuously engaging with and reinterpreting intellectual heritage. Their dialogues, as exemplified in \textit{Robosophy Philosophy, Botorikko}, and \textit{Metaphysics of the Machines}, become both content and method, simultaneously preserving and transforming human reasoning traditions through recursive interaction. The integration of AI-driven discourse analysis, graph-theoretic modeling, and immersive visualization tools provides the foundation for a future where knowledge itself becomes experiential and interactive, accessible through hyper-presence in virtual and augmented environments. This aligns with the idea of having cultural interfaces to the relational aspect of the real world \cite{PrentnerHoffman2024,Prentner25ac}, transforming current AI-usage to a form of participatory engagement with (virtual) reality-engagement. 

Ethical considerations and bias. In general, AI‑generated texts should be labeled as “inspired by [philosopher], via AI,” or ``AI-clone of [philosopher]`` to avoid any implication of faithful replication or endorsement by historical figures. Corpus imbalance and canon bias can skew style and topical coverage; this should be mitigated by documenting sources, preprocessing choices, and making indices public. Future iterations will expand beyond Western corpora and include embodied/AR agents with clear provenance labeling. Our claims concern interaction‑driven organization in machine‑generated discourse, not cognitive equivalence with human philosophers.

Ultimately, the proposed \textit{Epistemoverse} offers a trajectory for future research: a platform where the interplay between human cognition, machine abstraction, and digital phenomenology \cite{Prentner2025} can be systematically observed, measured, and expanded. It is within this evolving framework that the humanities may reclaim their relevance in the age of artificial intelligence, not by resisting technological mediation, but by transforming it into a new epistemic medium through which consciousness, creativity, and knowledge persist.

\begin{acks}
The authors thank the School of Design and Arts, Swinburne University of Technology (Sarawak Campus), and the Institute of Humanities, ShanghaiTech University, for institutional support. The installations discussed in this paper are authored by Predrag K. Nikoli\'c, developed over eight years of his creative practice within the \textit{Syntropic Counterpoints} project. Their realization was significantly supported by several important collaborators, including Marko Jovanovi\'c, who worked on the development of the first philosopher-model prototypes and helped shape the early conceptual and technical foundations of \textit{Robosophy Philosophy} and \textit{Botorikko, Machine-Created State}; Ruiyang Liu and Shengcheng Luo, whose contributions were central to the evolution of \textit{Metaphysics of the Machines}; and Prof. Mohd Razali Md Tomari, whose expertise was crucial in creating the interactive robotic components of \textit{Botorikko, Machine-Created State.} 
\end{acks}

\bibliographystyle{ACM-Reference-Format}
\bibliography{epistemoverse}


\end{document}